\documentclass[lettersize,journal]{IEEEtran}
\usepackage{amsmath,amsfonts}
\usepackage{algorithmic}
\usepackage{algorithm}
\usepackage{array}
\usepackage[caption=false,font=normalsize,labelfont=sf,textfont=sf]{subfig}
\usepackage{textcomp}
\usepackage{stfloats}
\usepackage{url}
\usepackage{verbatim}
\usepackage{graphicx}
\usepackage{cite}
\usepackage{multirow} 
\usepackage{color}

\begin{document}

\title{SAWU-Net: Spatial Attention Weighted Unmixing Network for Hyperspectral Images}

\author{IEEE Publication Technology,~\IEEEmembership{Staff,~IEEE,}
        % <-this % stops a space
\thanks{This paper was produced by the IEEE Publication Technology Group. They are in Piscataway, NJ.}% <-this % stops a space
\thanks{Manuscript received April 19, 2021; revised August 16, 2021.}}
\author{Lin~Qi,
        Xuewen~Qin,        
        Feng~Gao,~\IEEEmembership{Member,~IEEE},
        Junyu~Dong,~\IEEEmembership{Member,~IEEE},
        and
        Xinbo~Gao,~\IEEEmembership{Senior Member,~IEEE}

\thanks{This work was supported in part by the National Natural Science Foundation of China under Grant 42106191 and Grant U22A2096, and in part by the China Postdoctoral Science Foundation under Grant 2021M693023 and Grant 2022T150608. (Lin Qi and Xuewen Qin contribute equally to this work. Corresponding author: Feng Gao.)}
\thanks{Lin Qi, Xuewen~Qin, Feng Gao and Junyu Dong are with the School of Computer Science and Technology, Ocean University of China, Qingdao 266100, China (e-mail: gaofeng@ouc.edu.cn).}
\thanks{Xinbo Gao is with the Chongqing Key Laboratory of Image Cognition, Chongqing University of Posts and Telecommunications, Chongqing 400065, China (e-mail: gaoxb@cqupt.edu.cn).}

}

\maketitle

\begin{abstract}
Hyperspectral unmixing is a critical yet challenging task in hyperspectral image interpretation. Recently, great efforts have been made to solve the hyperspectral unmixing task via deep autoencoders. 
However, existing networks mainly focus on extracting spectral features from mixed pixels, and the employment of spatial feature prior knowledge is still insufficient.
To this end, we put forward a spatial attention weighted unmixing network, dubbed as SAWU-Net, which learns a spatial attention network and a weighted unmixing network in an end-to-end manner for better spatial feature exploitation.
In particular, we design a spatial attention module, which consists of a pixel attention block and a window attention block to efficiently model pixel-based spectral information and patch-based spatial information, respectively. While in the weighted unmixing framework, the central pixel abundance is dynamically weighted by the coarse-grained abundances of surrounding pixels.
In addition, SAWU-Net generates dynamically adaptive spatial weights through the spatial attention mechanism, so as to dynamically integrate surrounding pixels more effectively.
Experimental results on real and synthetic datasets demonstrate the better accuracy and superiority of SAWU-Net, which reflects the effectiveness of the proposed spatial attention mechanism.

\end{abstract}

\section{Introduction}
\IEEEPARstart{R}{ecent} years have witnessed the rapid growth of hyperspectral images (HSIs), and HSIs have been widely used in a broad range of applications due to the abundant spectral information  \cite{bioucas2012hyperspectral, 9617619,7508992}. However, the spatial resolution of HSI is sometimes limited, and thus the mixed pixel problem arises \cite{bioucas2012hyperspectral}, \cite{7508992}. Hyperspectral unmixing (HU) aims to decompose mixed pixels into a collection of spectral signatures (endmembers) and a set of fractional abundances. Spectral mixing models can be characterized as linear or nonlinear \cite{bioucas2012hyperspectral}. In this work, we mainly focus on linear spectral unmixing, which is more popular since its simple model structure and lower computation burden. The prophase LMM is able to separate into geometrical, statistical, and sparse regression problems.

The major representative methods for geometry-based HU is the vertex component analysis (VCA) \cite{nascimento2005vertex}. Furthermore, when the spectral signals of HSI are extremely mixed, statistical-based methods become a valid choice \cite{qian2011hyperspectral}. Sparse regression-based unmixing is also an extremely dynamic research area, and a large number of unmixing algorithms based on sparse regression have emerged such as the collaborative sparse unmixing algorithm \cite{WeiTang2015SparseUO}.
Considering that spatial information is also vital for HU, a few NMF and sparse regression-based variants incorporating spatial information have been proposed \cite{zhu2014spectral}, \cite{iordache2012total}.

In recent years, deep learning has been extensively applied in HU, and deep autoencoder (AE) network is the mainstream network structure.
The encoder of AE generates the abundance scores, while the weights of the decoder are taken as endmembers.
Among them, EndNet\cite{ozkan2019endnet}, CNNAEU \cite{palsson2021convolutional}, TANet\cite{jin2021tanet}, SSAE\cite{huang2020spatial}, SSCU-Net\cite{9709843} are typical unmixing networks based on AE structure.
CNNAEU uses patch-based convolution operation to exploit spatial information.
While SSAE uses fixed weights generated by the spatial mathematical model. 
Meanwhile, SSCU-Net introduces a superpixel segmentation method based on abundance information, which considerably facilitates the use of spatial information. 
On the other hand, there are also relevant approaches that apply attention mechanism to explore the spatial information contained in HSIs.
MUNet \cite{9724217} applies attention mechanism to LiDAR data to guide hyperspectral image unmixing, and \cite{2020Attention} introduces attention mechanism to supervised hyperspectral image unmixing.

\begin{figure*}[ht]
\centering
\includegraphics[scale=0.55]{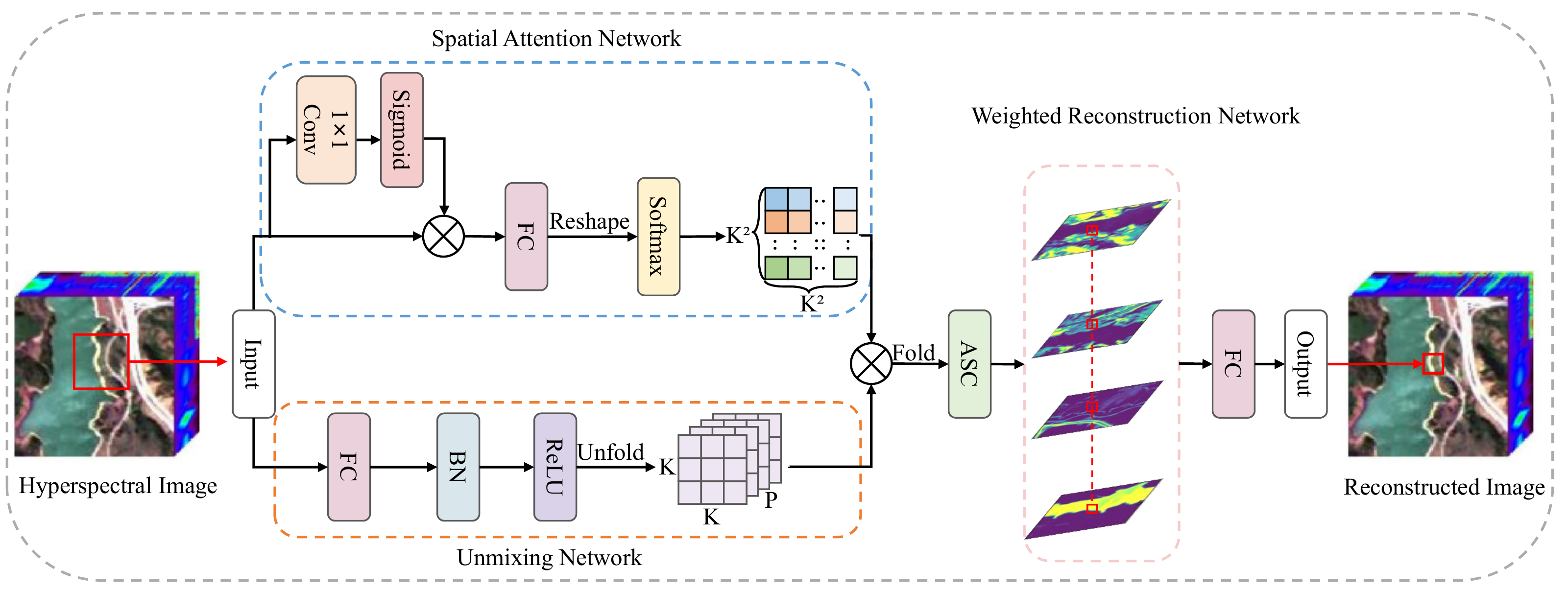}
\caption{ Flowchart of the proposed SAWU-Net architecture.}
\label{fig:pathdemo}
\end{figure*}

To sum up the above assay, introducing spatial information into the unmixing process and making full use of the spatial-spectral information is an essential topic for the HU task. However, the aforementioned utilization of spatial information is based on spatial mathematical models, which are often static and modeled based on specific assumptions and parameters. Therefore, to achieve dynamic adaptive spatial weights, in this paper, we propose a novel spatial attention weighted unmixing network, called SAWU-Net. 
Different from the attention mentioned above, our main motivation is to achieve dynamically adaptive spatial weights through spatial attention mechanism. It can continuously adjust the weight distribution according to the continuous learning of the network to generate more effective spatial weights. The main contributions of this paper can be summarized in the following three aspects.

\begin{itemize}
    \item[1)]
    In order to fully exploit the spatial information, we consider the contribution of surrounding pixels to the central pixel, thus introducing spatial attention to achieve the dynamic weight distribution of spatial information, and then propose a novel general spatial attention weighted unmixing network. SAWU-Net is a simple and efficient two-branch structure, where the weighted unmixing network is efficiently trained in an end-to-end manner to estimate endmembers and abundances simultaneously.
    
    \item[2)]
    In particular, we design a spatial attention module that consists of two parts. The first part is the pixel attention block, which efficiently interprets and enhances spectral information through 1×1 convolution. While the second part is the window attention block, which generates more effective spatial attention weights through a linear projection with low computational intensity.
    
    \item[3)]

    Moreover, in the field of hyperspectral unmixing, this is the first unmixing network that uses spatial attention to dynamically generate spatial weights. We introduce the spatial attention mechanism on the basic and simple AE network, which can fully explore the contribution of the spatial attention mechanism. Through experiments on real and synthetic datasets, the effectiveness and superiority of SAWU-Net are firmly demonstrated.

\end{itemize}

\section{Proposed Framework}

To sufficiently and effectively exploit the spatial information in HSIs, we propose a novel spatial attention weighted unmixing network.
Fig. 1 shows our proposed SAWU-Net architecture.
The network consists of two parts, including an attention network that generates a weight matrix of spatial information, and a simple three-layer AE unmixing network. In the following, we elaborate on the proposed SAWU-Net architecture.

\subsection{Spatial Attention Network}
In SAWU-Net, a local window of size $K \times K$ and L bands centered on the $x_{i,j}$ pixel is selected.  Let $x_{\Delta{i,j}}$ denote all the pixels within the local window centered at $(i, j), i.e.$,

\begin{equation}
    x_{\Delta_{i,j}} =\Big \{ {{x_{i+p-\lfloor \frac{K}{2} \rfloor,j+q-\lfloor \frac{K}{2} \rfloor}} }\Big \},0\leq p,q<K
\end{equation}

Then the abundance of $x_{i,j}$ is obtained by spatial attention weighting of the abundance of $x_{\Delta{i,j}}$, which is used to reconstruct $x_{i,j}$. 

Considering the characteristics of high spectral dimension and low spatial resolution of hyperspectral images, we effectively combine two well-known types of attention \cite{zhao2020efficient}, \cite{9888055},  and propose a novel attention module.
It concatenates two blocks, focusing on pixel-level and patch-level features, respectively. Pixel attention based on spectral information can assist window attention to obtain more effective dynamic spatial weights, and then obtain better unmixing results.

The pixel attention block obtains the pixel attention weight map through 1×1 convolution and sigmoid activation function, and then weighted to the input pixel spectral dimension.
It can assign greater weights to important spectral bands, while reduce the weights of redundant bands, thereby achieve the enhancement of spectral bands. In other words, the pixel attention block can capture or highlight certain important spectral bands. Pixel attention (PA) can be written as follows:

\begin{equation}
    x_{i,j}^{pa} = x_{i,j} \odot \sigma(Conv(x_{i,j}))
\end{equation}
where $Conv$ represents $1 \times 1$ convolution, $\sigma$ represents the sigmoid activation function, and $\odot$ indicates elementwise multiplication.

Meanwhile, a window attention block is connected in series with the pixel attention block. The purpose is to make better use of the spectral information features enhanced by pixel attention, and to generate more efficient spatial weights through effective linear projection, thereby significantly improving the performance of hyperspectral unmixing. Window attention can be written as follows:

\begin{equation}
\hat{D}_{i,j} = Softmax(Reshape(MLP(x_{i,j}^{pa})))
\end{equation}

\begin{figure*}[ht]
    \begin{center}
        \includegraphics[scale=0.52]{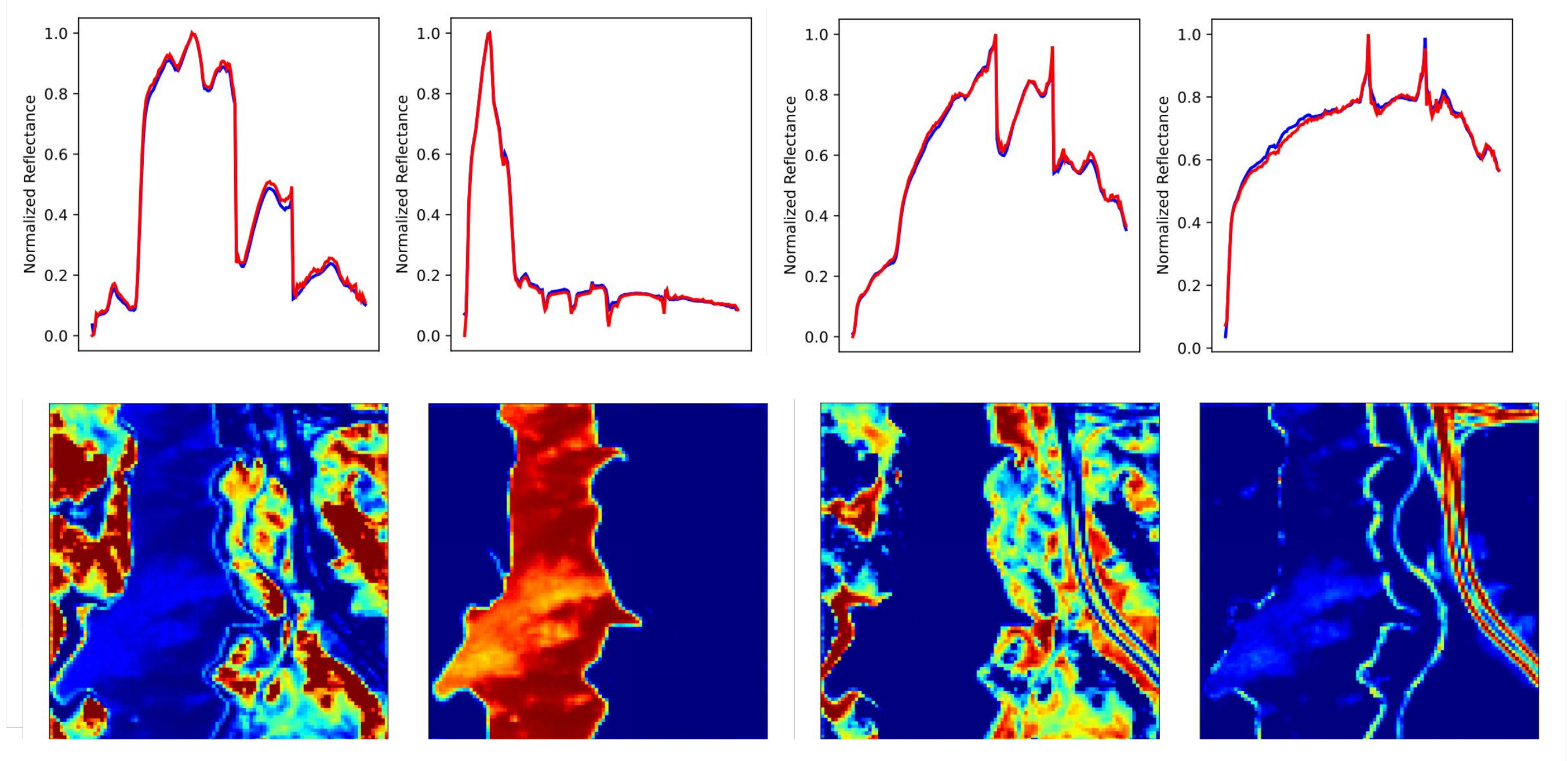}
        \caption{ Endmembers estimated by SAWU-Net (blue curves), the GTs (red curves) and the abundance maps on Jasper Ridge dataset.}
        \label{Em_Abundance}
    \end{center}
\end{figure*}

For the input, $x_{i,j}^{pa}$ through a  fully connected layer, the linear layers of weights  $W_D\in\mathbb{R}^{L\times K^4}$. Via the reshape operation, and then by a softmax activation function, an attention weight map is generated $\hat{D}_{i,j} \in\mathbb{R}^{K^2 \times K^2}$, for the weighting of abundance.

\subsection{Unmixing Network}

The unmixing network proposed in this paper is designed as a simple three-layer AE network, which is the same as common unmixing networks\cite{ozkan2019endnet}, \cite{9709843}. The encoder is designed as a fully connected layer to generate coarse-grained abundances, which are finally dynamically augmented by spatial attention weights.

The encoder mainly uses the ReLU activation function, in order to limit abundance nonnegativity constraint (ANC). Moreover, BN and Dropout denote the batch normalization layer and dropout layer, respectively. Therefore, the hidden representation obtained by fully connected layer encoding is as follows:

\begin{equation}
  \begin{aligned}
    h_{i,j} & = ReLU (Dropout (BN (W^{(e)}x_{i,j}))) ,\\
    x_{i,j} & \in x_{\Delta{i,j}} , i,j = 1, 2, ..., K
  \end{aligned}
\end{equation}
where $W^{(e)}$ is the weight coefficient of the fully connected layer without bias.

\subsection{Weighted Reconstruction Network}

The unmixing network has obtained coarse-grained abundance values, therefore, we define $h_{\Delta_{i,j}}\in\mathbb{R}^{P \times K^2}$ to represent all abundance values with $P$ endmember signatures and $K \times K $ window centered at $(i,j),i.e.,$

\begin{equation}
    h_{\Delta_{i,j}} =\Big \{ {{h_{i+p-\lfloor \frac{K}{2} \rfloor,j+q-\lfloor \frac{K}{2} \rfloor}} }\Big \},0\leq p,q<K
\end{equation}

Thus, fusing the spatial attention weights, the abundance value projection procedure can be written as:

\begin{equation}
    H_{\Delta_{i,j}}=MatMul(\hat{D}_{i,j} ,h_{\Delta_{i,j}})
\end{equation}

Finally, through the fold operation, the abundances from different positions of the current window are dynamically summed up as the abundance vector output of the center pixel, and the operation can be presented as follows:

\begin{equation}
    \hat{H}_{i,j} = \sum\limits_{0\leq m,n<K}H^{i,j}_{\Delta_{i+m-\lfloor \frac{K}{2} \rfloor,j+n-\lfloor \frac{K}{2} \rfloor}}
\end{equation}

Meanwhile, to satisfy the abundance sum-to-one constraint (ASC), we use the $l_1$ norm. Therefore, the eventual abundance of the central pixel can be expressed as:

\begin{equation}
    s_c=\frac{\hat{H}_{i,j}}{\Vert \hat{H}_{i,j}\Vert _1+\varepsilon}
\end{equation}
where $\varepsilon$ is a very small number to prevent the denominator from being meaningless.

Eventually, through the decoder, the reconstructed center pixel can be expressed as follows:

\begin{equation}
    \hat{x}_{i,j} =W^{(d)}s_c
\end{equation}
where $W^{ (d)}$ represents the weight matrix of the fully connected layer of the decoder, which corresponds to the extracted endmember matrix.

The weighted reconstruction network in the overall architecture can assign different weights to the input features according to their importance, which helps to enhance the performance of hyperspectral unmixing. In addition, using a simple network structure is also beneficial to tap the potential of SAWU-Net, exploring a unified network for unmixing, and embodying the effectiveness of spatial attention weights.

\subsection{Initialization and loss function}

Experiments have proved that the spectral angular distance (SAD) as a loss function of the AE unmixing network has obtained superior unmixing results. At the same time, in order to improve the sparsity of abundance, we impose a constraint with the $l_{1/2}$ norm. Therefore, the overall loss of SAWU-Net can be formulated as:

\begin{figure}[ht]
    \begin{center}
        \includegraphics[scale=0.4]{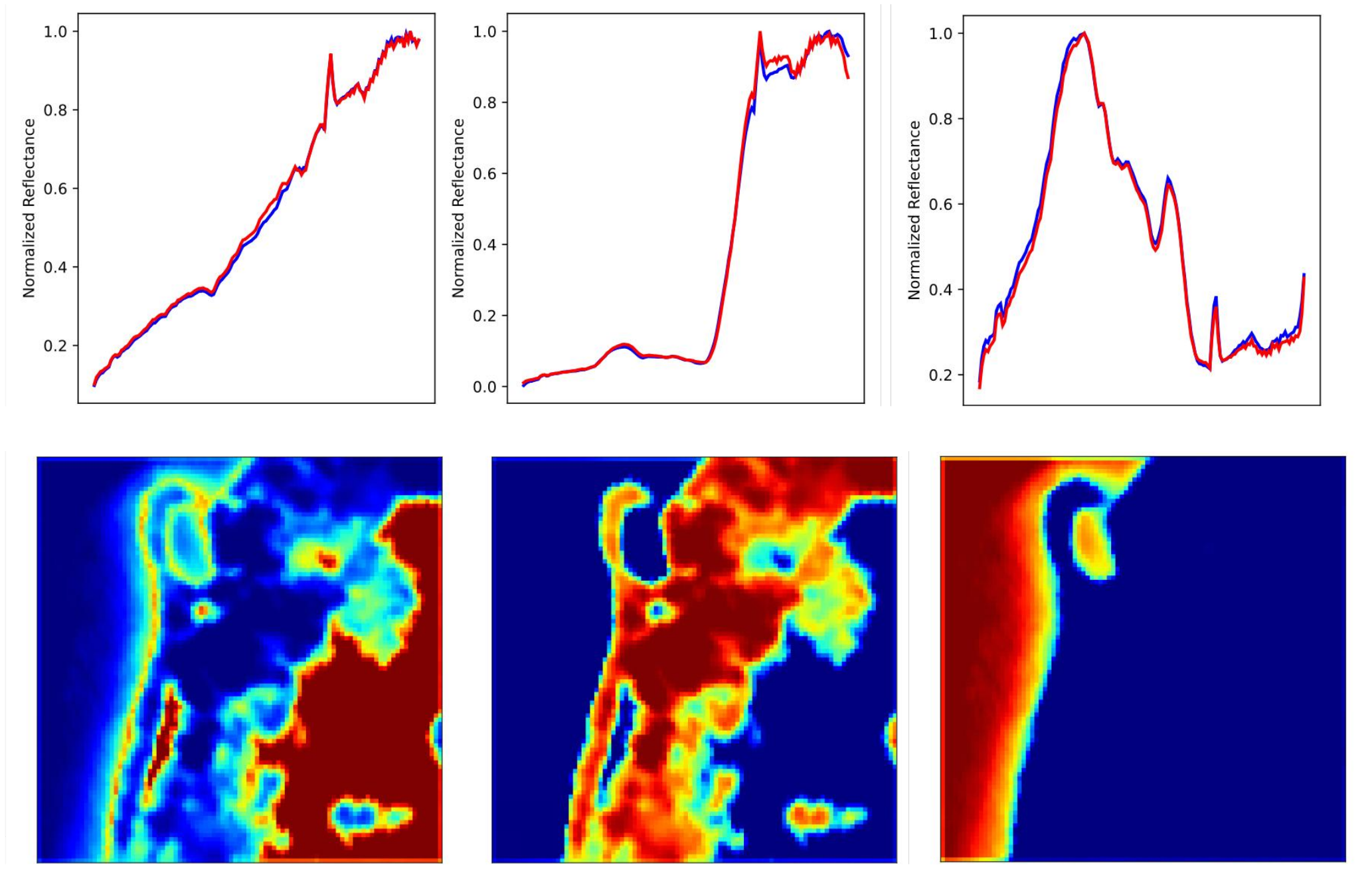}
        \caption{ Endmembers estimated by SAWU-Net (blue curves), the GTs (red curves) and the abundance maps on Samson dataset.}
        \label{Samson_Em_Abu}
    \end{center}
\end{figure}

\begin{table}[h!]
	\centering
        \caption{Hyperparameters Setting.}
        \begin{tabular}{l | cc}\hline\hline
      \multirow{2}*{Hyperparameters} &  \multicolumn{2}{c}{Datasets}\\
      & Jasper Rage & Samson  \\ \hline
      $\lambda_1$ & \multicolumn{2}{c}{12} \\
      $\lambda_2$ & \multicolumn{2}{c}{$2e$-$3$} \\
      Batch size & \multicolumn{2}{c}{128} \\
      Optimizer &  \multicolumn{2}{c}{Adam}   \\
      Maxiter &  \multicolumn{2}{c}{300}   \\
      Learning rate (encoder) & \multicolumn{2}{c}{$1e$-$3$ }\\
      Learning rate (decoder) &$1e$-$5$ & $1e$-$4$\\ \hline\hline
    \end{tabular}
	\label{table_res_san}
\end{table}

\begin{equation}
    L_{SAWU} = \lambda_1cos^{-1}\frac{\langle {x}_{i,j},\hat{x}_{i,j}\rangle}{\vert{x}_{i,j}\vert_2\vert \hat{x}_{i,j}\vert_2}+\lambda_2\vert s_c\vert_{\frac{1}{2}}
\end{equation}
where ${x}_{i,j}$ and $\hat{x}_{i,j}$ represent the original center pixel and the reconstructed center pixel, respectively, and $s_c$ represents the abundance vector of the center pixel after spatial attention weighting. Meanwhile $\lambda_1$ and $\lambda_2$ are the regularization parameters, where its specific values are shown in Table I.

\section{Experiments}
To further assess our proposed method, we compare with traditional and deep learning-based linear unmixing methods. They are VCA \cite{nascimento2005vertex}, L$_{1/2}$-NMF \cite{qian2011hyperspectral}, Dgs-NMF \cite{zhu2014spectral}, EndNet \cite{ozkan2019endnet}, CNNAEU \cite{palsson2021convolutional}, SSAE \cite{huang2020spatial}, respectively. EndNet only uses spectral information, while CNNAEU and SSAE make full use of spatial-spectral information.

\subsection{Experimental Setup and Data Description}

\begin{table}[h!]

    \tabcolsep=0.15cm
    \renewcommand\arraystretch{1.3} 
    \tiny
    
	\centering
	\caption{Valuation metrics SAD and RMSE for abundance and endmember of Jasper Ridge dataset. Best result are bold.}
    \begin{tabular}{m{0.1cm} ccccccc}\hline\hline
    
    \multicolumn{8}{c}{Spectral Angle Distance (×$10^{-2}$)}  \\

       EM & VCA & $L_{1/2}$-NMF & Dgs-NMF & EndNet & CNNAEU & SSAE & SAWU \\\hline

           \#1 & 13.95 ± 3.1 & 15.10 ± 0.3 & 4.66 ± 0.2 & 4.57 ± 0.4 & 11.94 ± 2.1 & 3.37 ± 0.1 &\textbf{2.63 ± 0.2}    \\
           \#2  & 29.09 ± 10.2 & 4.60 ± 0.0 & 4.60 ± 0.0 & 5.05 ± 0.9 & 6.92 ± 0.4 & 4.72 ± 0.1 &\textbf{3.03 ± 0.1} \\
           \#3 & 15.22 ± 2.3 & 6.16 ± 0.5 & 5.66 ± 0.2 & 5.29 ± 0.3 & 10.15 ± 0.9 & 3.02 ± 0.3 &\textbf{2.50 ± 0.1} \\
           \#4 & 9.94 ± 1.1 & 9.81 ± 0.1 & 6.73 ± 0.1 & 3.54 ± 0.2 & 7.45 ± 0.4 & 2.77 ± 0.2 &\textbf{1.68 ± 0.3}\\
           Avg & 17.05 ± 3.6 & 7.19 ± 2.4 & 5.41 ± 0.1 & 4.61 ± 0.5 & 9.12 ± 0.6 & 3.47 ± 0.1 &\textbf {2.46 ± 0.1} \\ \hline
          
          \multicolumn{8}{c}{Root Mean Square Error (×$10^{-2}$)}  \\
           EM & VCA & $L_{1/2}$-NMF & Dgs-NMF & EndNet & CNNAEU & SSAE & SAWU \\\hline
           \#1 & 16.09 ± 3.1 & 16.16 ± 0.5 & 11.66 ± 0.2 & 8.85 ± 0.4 & 13.56 ± 0.3 &\textbf{5.15 ± 0.6} & 5.54 ± 0.4\\
           \#2  & 6.06 ± 1.8 & 5.57 ± 0.0 & 4.13 ± 0.0 & 6.88 ± 0.3 & 9.66 ± 0.3 & 5.63 ± 0.4 & \textbf{4.01 ± 0.2} \\
           \#3 & 15.00 ± 1.5 & 17.02 ± 0.4 & 11.13 ± 0.3 & 10.59 ± 0.2 & 10.61 ± 0.2 &\textbf {6.19 ± 0.4} & 
           6.28 ± 0.1\\
           \#4 & 11.09 ± 1.5 & 6.73 ± 0.2 & 5.68 ± 0.1 & 11.17 ± 0.4  & 8.64 ± 0.2 & 7.15 ± 0.3 &\textbf{5.37 ± 0.3}\\
           Avg & 12.05 ± 1.5 & 11.37 ± 0.2 & 8.15 ± 0.2 & 9.37 ± 0.5 & 10.62 ± 0.1 & 6.03 ± 0.2 & \textbf{5.30 ± 0.2}\\ \hline \hline

    \end{tabular}
	\label{table_res_san}
\end{table}

\begin{table}[h!]

    \tabcolsep=0.15cm
    \renewcommand\arraystretch{1.3} 
    \tiny
	\centering
	\caption{Valuation metrics SAD and RMSE for abundance and endmember of Samson dataset. Best result are bold.}
    \begin{tabular}{m{0.1cm} ccccccc}\hline\hline
      \multicolumn{8}{c}{Spectral Angle Distance (×$10^{-2}$)}  \\

       EM & VCA & $L_{1/2}$-NMF & Dgs-NMF & EndNet & CNNAEU & SSAE & SAWU \\\hline
       \#1 & 4.21 ± 0.0 & 6.21 ± 7.3 & 5.64 ± 7.4 & 1.98 ± 0.2 & 3.73 ± 2.1 & 2.04 ± 0.1 &\textbf {1.42 ± 0.1}\\
       \#2  & 5.58 ± 0.2 & 5.23 ± 0.3 & 4.80 ± 0.3 & 5.31 ± 0.3 & 3.97 ± 0.4 & 3.61 ± 0.1 &\textbf {2.73 ± 0.2} \\
       \#3 & 43.97 ± 31.1 & 11.97 ± 2.1 & 4.70 ± 0.3 & 3.98 ± 0.2 & 4.30 ± 0.9 & 3.12 ± 0.2 &\textbf{2.30 ± 0.1} \\
       Avg & 17.92 ± 1.5 & 7.80 ± 3.2 & 5.05 ± 2.7 & 3.76 ± 0.2 & 4.00 ± 0.7 & 2.92 ± 0.1 &\textbf {2.15 ± 0.1}\\ \hline
  
      \multicolumn{8}{c}{Root Mean Square Error (×$10^{-2}$)}  \\
       EM & VCA & $L_{1/2}$-NMF & Dgs-NMF & EndNet & CNNAEU & SSAE & SAWU \\\hline
       \#1 & 16.45 ± 3.0 & 8.58 ± 3.3 & 7.77 ± 3.8 & 8.60 ± 0.0 & 18.49 ± 0.6 &\textbf {4.70 ± 0.4} & 5.66 ± 0.3\\
       \#2  & 11.25 ± 1.0 & 7.44 ± 3.7 & 7.74 ± 3.6 & 6.92 ± 0.1 & 16.21 ± 0.4 & 5.33 ± 0.3 & \textbf{5.31 ± 0.4} \\
       \#3 & 19.21 ± 3.3 & 5.55 ± 0.9 & 2.70 ± 0.9 & 4.99 ± 0.0 & 6.69 ± 0.3 & 5.27 ± 0.3 & \textbf{4.03 ± 0.3} \\
       Avg & 15.64 ± 0.6 & 7.19 ± 2.4 & 6.07 ± 2.8 & 6.84 ± 0.0 & 13.80 ± 0.4 & 5.10 ± 0.2 &\textbf{5.00 ± 0.2}\\ \hline\hline

    \end{tabular}
	\label{table_res_san}
\end{table}

\subsubsection{Evaluation Metrics}
So as to better assess the proposed network, we adopt the SAD, which is employed to calculate endmembers, and root mean square error (RMSE), which is utilized to measure abundance, two most commonly used evaluation metrics. They are defined as 
$SAD= cos^{-1}\frac{\langle x_i,\hat{x}_i\rangle}{\vert x_i\vert_2\vert \hat{x}_i\vert_2}$
and $RMSE = \sqrt{\frac{1}{N}\sum\limits^{N}_{i=1}(s_i-\hat{s}_i)^2}$. The smaller their corresponding values are, the excellent unmixing consequence is.

The proposed network is specifically implemented in the CPU environment using PyTorch and Python. The detailed network hyperparameters are shown in Table I.

\subsubsection{Hyperspectral Datasets}
We use two real datasets, Jasper Ridge and Samson, to comprehensively test the validity of the proposed method.
The datasets are processed in the same way as the references \cite{qian2011hyperspectral}, \cite{zhu2014spectral}. 
We also conduct experiments on a synthetic dataset, which is processed in the same way as reference \cite{WeiTang2015SparseUO}. Meanwhile, the SNR is set to 30 dB.

\subsection{Comparison of SAWU-Net Against Other Methods}

The experimental results of SAD and RMSE are obtained after 20 experiments, and the average value is secured.

\subsubsection{Experiments on Jasper Ridge DataSet}
Table II exhaustively lists the endmember extraction results for all compared unmixing methods. Compared with traditional unmixing algorithms, deep learning-based methods show strong performance. However, SAWU-Net utilizes spatial attention weighting mechanism to adaptively adjust the information weight of different spatial positions, which is beneficial for accurately extracting endmember information. At the same time, Fig. 2 shows the difference between the endmembers obtained by the proposed method and the ground truth (GT).

\begin{figure}[ht]
    \begin{center}
        \includegraphics[scale=0.31]{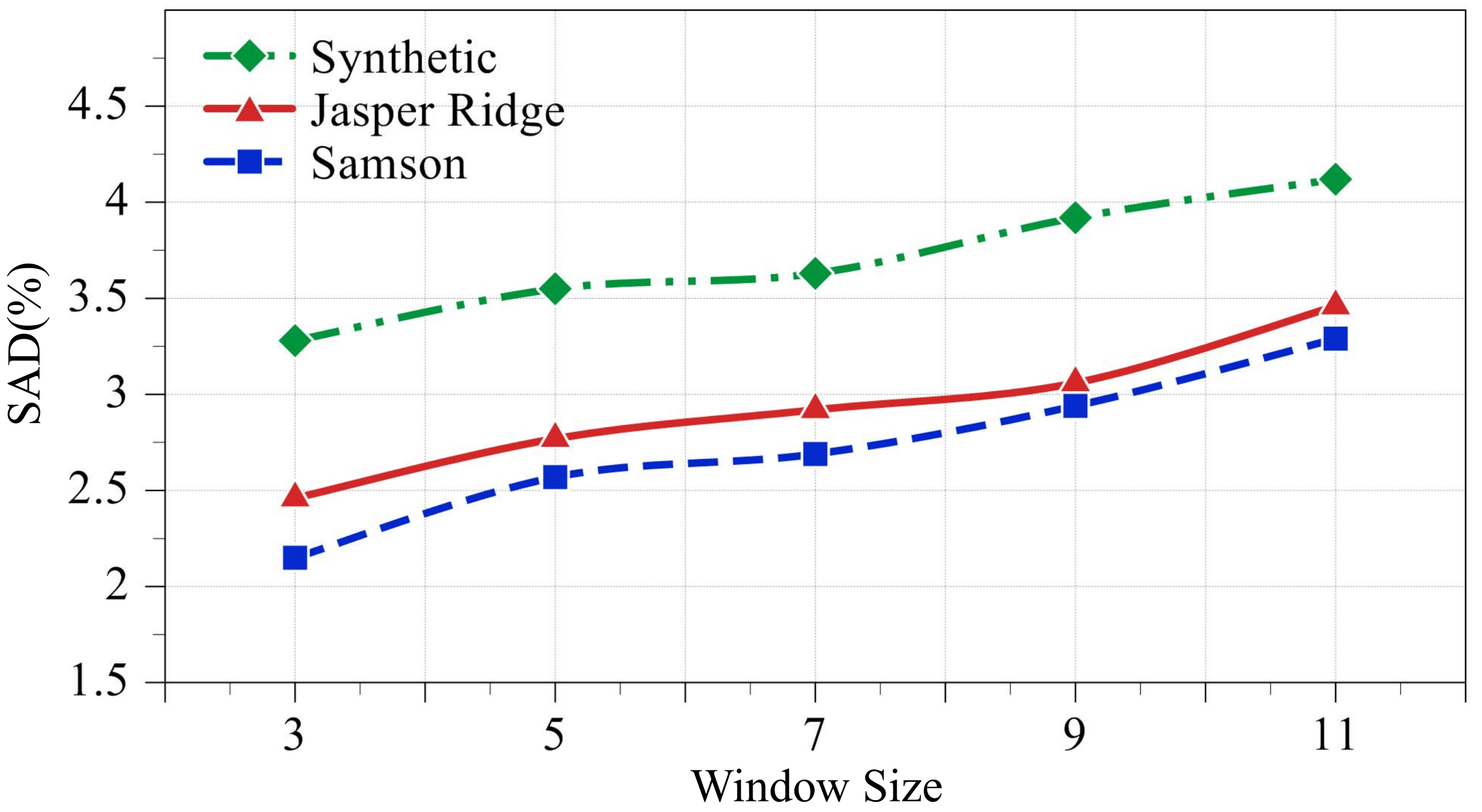}
        \caption{Quantitative analysis of sliding window size.}
        \label{Window_size}
    \end{center}
\end{figure}

Table II shows the results of abundance estimation on the Jasper Ridge dataset. Since the input to our network is window size data, in order to generate abundance maps for the entire hyperspectral image, we perform certain processing on the input data such that each pixel contains window size data, which is then fed into the network to yield our final abundances. From Table II, SAWU-Net effectively performs weighted fusion by considering the different contributions of surrounding pixels to the central pixel, so as to obtain more accurate abundance. At the same time, Fig. 2 visualizes the abundance map more intuitively.

\subsubsection{Experiments on Samson DataSet}
Table III detail the experimental results of the proposed and other comparative algorithms on the Samson dataset. Obviously, it performs about the same as the Jasper dataset. One possible reason is that the two datasets are relatively similar. As expected, the spatial attention weights obtained by SAWU-Net using spectral information are crucial for the extraction of endmembers.

In the abundance estimation stage, we still use the same processing method as the Jasper dataset. This is shown in Table III. Although the effect is comparable compared to SSAE, it still has a great advantage over other algorithms which shows that the SAWU-Net framework is versatile. Fig. 3 shows in detail the comparison between the endmembers obtained by the SAWU-Net and GTs, and visualizes the abundance map generated by the SAWU-Net.

\subsection{Ablation Study}

In our proposed SAWU-Net network, spatial attention weights serve as a key element. Whether to consider it or not has become the main research direction of our ablation experiment. At the same time, the window size has a certain influence on the endmember extraction.

\subsubsection{Experiments on window size}
In order to explore the correlation between the center pixel and the surrounding pixels, we selected five window sizes for experiments and finally found that the 3×3 window size is the most suitable for endmember extraction. Fig. 4 shows the experimental results in detail.

\subsubsection{Experiments on spatial attention}
Our ablation experiments mainly compare the most basic three-layer AE network and the EENet network in which the spatial weights are obtained mathematically in the SSAE network. Meanwhile, the column of SAWU-Net (w/o PA) represents the unmixing effect without the pixel attention block. Table IV shows that SAWU-Net can extract endmembers more effectively, which also demonstrates the effectiveness of our proposed spatial attention module.

\begin{table}[h!]
    \tabcolsep=0.15cm
    \renewcommand\arraystretch{1.3} 
    \scriptsize
	\centering
    \caption{ABLATION STUDAY FOR SAD (×$10^{-2}$) TO EVALUATE THE BASELINE NETWORKS AND SAWU-NET.}
    \begin{tabular}{c|cccc}\hline\hline
       Datasets & Baseline & SSAE (EENet) & SAWU-Net (w/o PA) & SAWU-Net \\\hline
       Jasper Ridge & 4.51  & 3.47 & 2.58 & \textbf{2.46}   \\
       Samson   & 3.55 & 2.92 & 2.27 & \textbf{2.15} \\ 
       Synthetic   & 4.83 & 3.84 & 3.36 & \textbf{3.28} \\ \hline\hline

    \end{tabular}
	\label{table_res_san}
\end{table}

\section{Conclusion}
In this letter, in order to make full use of the spatial information, we propose a novel spatial attention weighted network structure, which is simple and effective. First, we design a spatial attention module, which consists of a pixel attention block and a window attention block.  
In addition, weighted unmixing network employs dynamically weighted neighborhood coarse-grained abundance method, which further promotes the spatial continuity of abundance. Compared with state-of-the-art methods, SAWU-Net achieves excellent performance on real and synthetic hyperspectral datasets.

\bibliographystyle{IEEEtran}

\bibliography{main.bib}

% Generated by IEEEtran.bst, version: 1.14 (2015/08/26)
\begin{thebibliography}{10}
\providecommand{\url}[1]{#1}
\csname url@samestyle\endcsname
\providecommand{\newblock}{\relax}
\providecommand{\bibinfo}[2]{#2}
\providecommand{\BIBentrySTDinterwordspacing}{\spaceskip=0pt\relax}
\providecommand{\BIBentryALTinterwordstretchfactor}{4}
\providecommand{\BIBentryALTinterwordspacing}{\spaceskip=\fontdimen2\font plus
\BIBentryALTinterwordstretchfactor\fontdimen3\font minus
  \fontdimen4\font\relax}
\providecommand{\BIBforeignlanguage}[2]{{%
\expandafter\ifx\csname l@#1\endcsname\relax
\typeout{** WARNING: IEEEtran.bst: No hyphenation pattern has been}%
\typeout{** loaded for the language `#1'. Using the pattern for}%
\typeout{** the default language instead.}%
\else
\language=\csname l@#1\endcsname
\fi
#2}}
\providecommand{\BIBdecl}{\relax}
\BIBdecl

\bibitem{bioucas2012hyperspectral}
J.~M. Bioucas-Dias, A.~Plaza, N.~Dobigeon, M.~Parente, Q.~Du, P.~Gader, and
  J.~Chanussot, ``Hyperspectral unmixing overview: Geometrical, statistical,
  and sparse regression-based approaches,'' \emph{IEEE J. Sel. Topics Appl.
  Earth Observ. Remote Sens.}, vol.~5, no.~2, pp. 354--379, Apr. 2012.

\bibitem{9617619}
F.~Luo, Z.~Zou, J.~Liu, and Z.~Lin, ``Dimensionality reduction and
  classification of hyperspectral image via multistructure unified
  discriminative embedding,'' \emph{IEEE Trans. Geosci. Remote Sens.}, vol.~60,
  pp. 1--16, 2022.

\bibitem{7508992}
F.~Luo, H.~Huang, Z.~Ma, and J.~Liu, ``Semisupervised sparse manifold
  discriminative analysis for feature extraction of hyperspectral images,''
  \emph{IEEE Trans. Geosci. Remote Sens.}, vol.~54, no.~10, pp. 6197--6211,
  2016.

\bibitem{nascimento2005vertex}
J.~M. Nascimento and J.~M. Bioucas-Dias, ``Vertex component analysis: A fast
  algorithm to unmix hyperspectral data,'' \emph{IEEE Trans. Geosci. Remote
  Sens.}, vol.~43, no.~4, pp. 898--910, Apr. 2005.

\bibitem{qian2011hyperspectral}
Y.~Qian, S.~Jia, J.~Zhou, and A.~Robles-Kelly, ``Hyperspectral unmixing via
  ${L_{1/2}}$ sparsity-constrained nonnegative matrix factorization,''
  \emph{IEEE Trans. Geosci. Remote Sens.}, vol.~49, no.~11, pp. 4282--4297,
  Nov. 2011.

\bibitem{WeiTang2015SparseUO}
W.~Tang, Z.~Shi, Y.~Wu, and C.~Zhang, ``Sparse unmixing of hyperspectral data
  using spectral a priori information,'' \emph{IEEE Trans. Geosci. Remote
  Sens.}, vol. 53, no. 2, pp. 770–783, Feb. 2015.

\bibitem{zhu2014spectral}
F.~Zhu, Y.~Wang, B.~Fan, S.~Xiang, G.~Meng, and C.~Pan, ``Spectral unmixing via
  data-guided sparsity,'' \emph{IEEE Trans. Image Process.}, vol.~23, no.~12,
  pp. 5412--5427, Dec. 2014.

\bibitem{iordache2012total}
M.-D. Iordache, J.~M. Bioucas-Dias, and A.~Plaza, ``Total variation spatial
  regularization for sparse hyperspectral unmixing,'' \emph{IEEE Trans. Geosci.
  Remote Sens.}, vol.~50, no.~11, pp. 4484--4502, Nov. 2012.

\bibitem{ozkan2019endnet}
S.~Ozkan, B.~Kaya, and G.~B. Akar, ``Endnet: Sparse autoencoder network for
  endmember extraction and hyperspectral unmixing,'' \emph{IEEE Trans. Geosci.
  Remote Sens.}, vol.~57, no.~1, pp. 482--496, Jan. 2019.

\bibitem{palsson2021convolutional}
B.~Palsson, M.~O. Ulfarsson, and J.~R. Sveinsson, ``Convolutional autoencoder
  for spectral--spatial hyperspectral unmixing,'' \emph{IEEE Trans. Geosci.
  Remote Sens.}, vol.~59, no.~1, pp. 535--549, Jan. 2021.

\bibitem{jin2021tanet}
Q.~Jin, Y.~Ma, X.~Mei, and J.~Ma, ``Tanet: An unsupervised two-stream
  autoencoder network for hyperspectral unmixing,'' \emph{IEEE Trans. Geosci.
  Remote Sens.}, vol. early access, 2021.

\bibitem{huang2020spatial}
Y.~Huang, J.~Li, L.~Qi, Y.~Wang, and X.~Gao, ``Spatial-spectral autoencoder
  networks for hyperspectral unmixing,'' in \emph{Proc. IEEE Int. Geosci.
  Remote Sens. Symp. (IGARSS'20)}, Sep. 2020, pp. 2396--2399.

\bibitem{9709843}
L.~Qi, F.~Gao, J.~Dong, X.~Gao, and Q.~Du, ``Sscu-net: Spatial–spectral
  collaborative unmixing network for hyperspectral images,'' \emph{IEEE Trans.
  Geosci. Remote Sens.}, vol.~60, pp. 1--15, 2022.

\bibitem{9724217}
Z.~Han, D.~Hong, L.~Gao, J.~Yao, B.~Zhang, and J.~Chanussot, ``Multimodal
  hyperspectral unmixing: Insights from attention networks,'' \emph{IEEE Trans.
  Geosci. Remote Sens.}, vol.~60, pp. 1--13, 2022.

\bibitem{2020Attention}
``Attention-based residual network with scattering transform features for
  hyperspectral unmixing with limited training samples,'' \emph{International
  journal of applied mechanics}, vol.~12, no.~3, 2020.

\bibitem{zhao2020efficient}
H.~Zhao, X.~Kong, J.~He, Y.~Qiao, and C.~Dong, ``Efficient image
  super-resolution using pixel attention,'' in \emph{European Conference on
  Computer Vision}, 2020, pp. 56--72.

\bibitem{9888055}
L.~Yuan, Q.~Hou, Z.~Jiang, J.~Feng, and S.~Yan, ``Volo: Vision outlooker for
  visual recognition,'' \emph{IEEE Trans. Pattern Anal. Mach. Intell.}, pp.
  1--13, 2022.

\end{thebibliography}

\end{document}